\documentclass[a4paper]{spie}  

\usepackage{amsmath,amsfonts,amssymb}
\usepackage{graphicx}
\usepackage[colorlinks=true, allcolors=blue]{hyperref}

\title{Curved detectors developments and characterization: application to astronomical instruments}
\author{Simona Lombardo \supit{a}, Thibault Behaghel \supit{a}, Bertrand Chambion \supit{c}, Wilfried Jahn \supit{b}, Emmanuel Hugot \supit{a}, Eduard Muslimov \supit{a}, Melanie Roulet \supit{a}, Marc Ferrari \supit{a}, Christophe Gaschet \supit{c}, St\'ephane Caplet \supit{c}, David Henry \supit{c} 
\skiplinehalf
\supit{a} Aix Marseille Univ, CNRS, CNES, LAM, Marseille, France; \\
\supit{b} Division of aerospace engineering, Caltech, Pasadena, CA 91125, USA;\\
\supit{c} Univ. Grenoble Alpes, CEA, LETI, MINATEC campus, F38054 Grenoble, France\\
}

\authorinfo{simona.lombardo@lam.fr}
\begin{document} 
\maketitle

\begin{abstract}
Many astronomical optical systems have the disadvantage of generating curved focal planes requiring flattening optical elements to project the corrected image on flat detectors. 
The use of these designs in combination with a classical flat sensor implies an overall degradation of throughput and system performances to obtain the proper corrected image.
With the recent development of curved sensor this can be avoided.
This new technology has been gathering more and more attention from a very broad community, as the potential applications are multiple: from low-cost commercial to high impact scientific systems, to mass-market and on board cameras, defense and security, and astronomical community. 

We describe here the first concave curved CMOS detector developed within a collaboration between CNRS-LAM and CEA-LETI.
This fully-functional detector 20\,Mpix (CMOSIS CMV20000) has been curved down to a radius of $R_\mathrm{c}=$150\,mm over a size of 24x32\,mm$^2$. 
We present here the methodology adopted for its characterization and describe in detail all the results obtained. 
We also discuss the main components of noise, such as the readout noise, the fixed pattern noise and the dark current. Finally we provide a comparison with the flat version of the same sensor in order to establish the impact of the curving process on the main characteristics of the sensor.

\end{abstract}


\section{INTRODUCTION}
\label{sec:intro}  
Many astronomical optical systems have the disadvantage of generating curved focal planes requiring flattening optical elements to project the corrected image on flat detectors. 
The use of these designs in combination with a classical flat sensor implies an overall degradation of throughput and system performances to obtain the proper corrected image.
This is extremely harmful in case the science goal requires an instrumental PSF as compact as possible e.g. to measure the ultra-low surface brightness universe.
One example for this is the space mission MESSIER which aims at measuring surface brightness levels as low as 35 mag arcsec$^{-2}$ in the optical (350-1000 nm) and 38 mag arcsec$^{-2}$ in the UV (200 nm).
Any refractive surface must also be excluded in this design as they would generate Cherenkov emission due to the relativistic particles.
This automatically eliminates the possibility of using optics to flatten curved focal plane.

With the recent development of curved detectors\cite{guenter2017,iwert_2012,dumas_2012,tekaya_2013,andata}, this is not an issue anymore. 
The introduction of curved detectors has two additional advantages: it allows to have a reduction of the size of the imaging systems, consequently making their design more compact and simpler (particularly important for space missions), and, at the same time, it improves their overall performances (e.g. higher resolution). 
This new technology has been gathering more and more attention from a very broad community, as the potential applications are multiple: from low-cost commercial to high impact scientific systems, to mass-market and on board cameras,defense and security \cite{tekaya_2014}, and astronomical community. 

A version of MESSIER with reduced dimensions is planned as its ground-based demonstrator \cite{eduard2017}.
This fully reflective Schmidt design will be installed in Tenerife and it will potentially be the first astronomical telescope that uses a curved detector -- with convex shape and a radius of curvature of 800\,mm -- probing the full capabilities of this groundbreaking technology.

In this paper, we describe the first concave curved CMOS detector developed within a collaboration between CNRS-LAM and CEA-LETI.
This fully-functional detector of 20\,Mpix (CMOSIS, CMV20000 \cite{cmv20000}) has been curved down to a radius of 150\,mm over a size of 24x32\,mm$^2$ and it is sensitive to the visible light. Specific care was taken to keep the packaging identical to the original one before curving, in such a way that the final product is a “plug-and-play” commercial component.

We present here the methodology adopted for its characterization (Section~\ref{sec:methodology}) and describe in detail all the results obtained (Section~\ref{sec:results}). 
We also discuss the main components of noise, such as the readout noise, the fixed pattern noise and the dark current. Finally we provide a comparison with the flat version of the same sensor in order to establish the impact of the curving process on the main characteristics of the sensor.

\section{CURVING PROCESS}
\label{sec:sections}

CEA-LETI in collaboration with CNRS-LAM is in a phase of prototyping several curved detectors concepts.
Such detectors have already shown promising results and demonstrated some of the improvements achievable in term of compactness and performances of the related optical designs \cite{cea_2018}. 

\begin{figure}[ht]
 \begin{center}
  \includegraphics[width=0.9\textwidth]{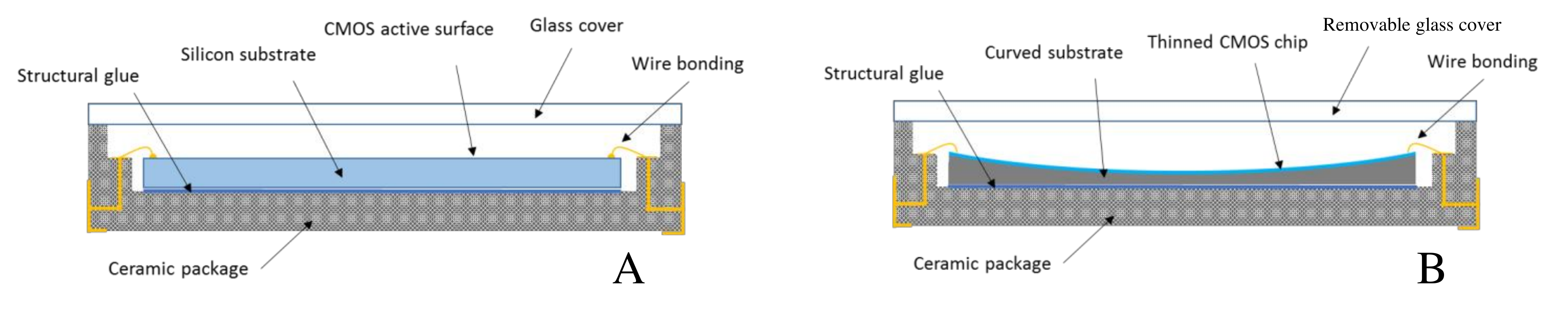}
  \caption{Schematic view of the commercial flat CMOS (A) and curved (B). From \cite{cea_2018}.}
  \label{cmos_struct}
 \end{center}
\end{figure}

The off-the-shelf initial flat sensor consists of a silicon die glued on a ceramic package.
Wire bonding from the die to the package surface provides the electrical connections. 
Finally a glass window protects the sensor surface from mechanical or environment solicitations (figure \ref{cmos_struct}A). 
The curving process of these sensors consists of two steps: firstly the sensors are thinned  with a grinding equipment to increase their mechanical flexibility, then they are glued onto a curved substrate. 
The required shape of the CMOS is, hence, due to the shape of the substrate.
The sensors are then wire bonded keeping the packaging identical to the original one before curving.
The final product is, therefore, a “plug-and-play” commercial component ready to be used or tested (figure \ref{cmos_struct}B).

\begin{figure}[ht]
 \begin{center}
  \includegraphics[width=0.5\textwidth]{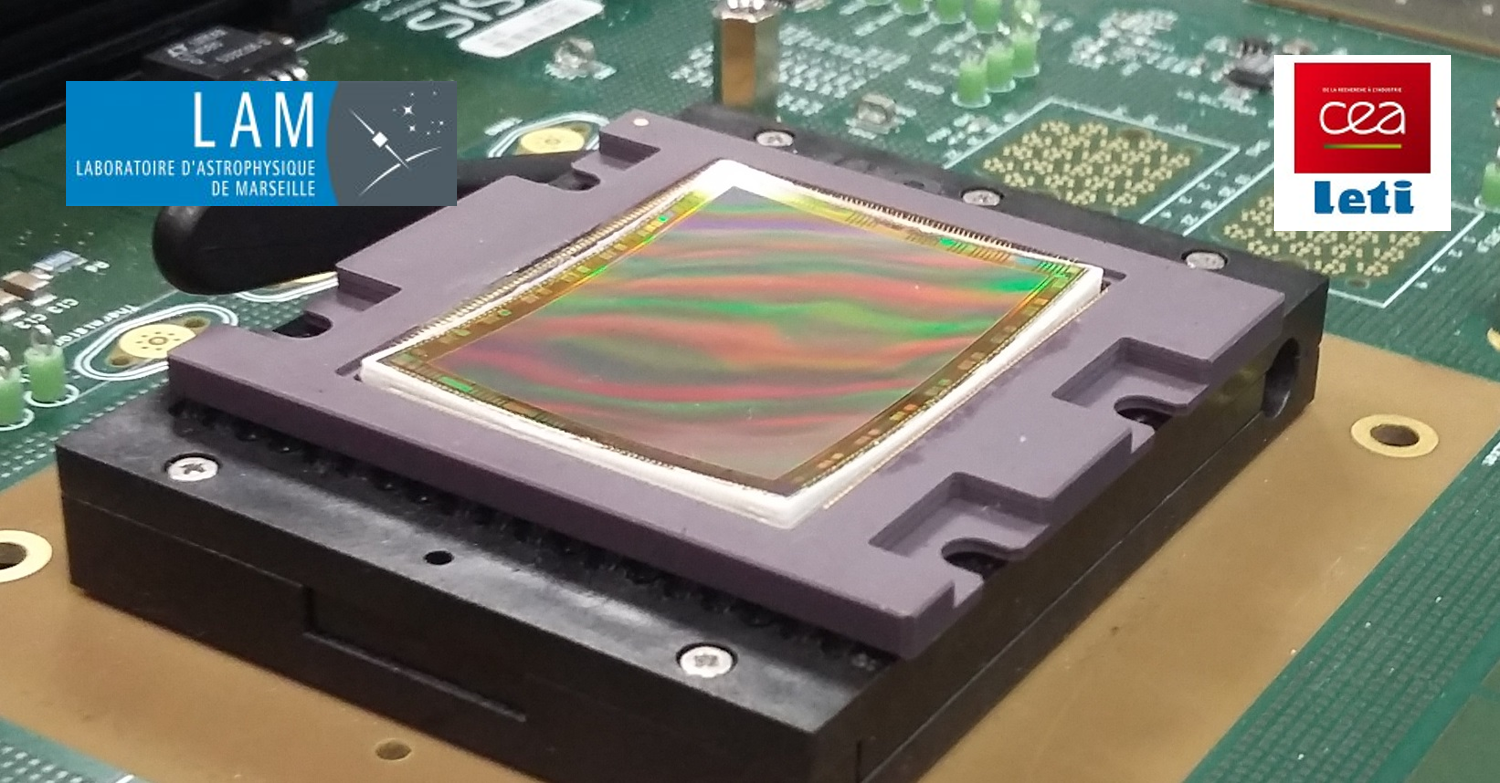}
  \caption{CMV20000 \cite{cmv20000} CMOS image sensor from CMOSIS plugged on the CMOSIS evaluation board. This sensor has been curved with a spherical concave shape down to a radius of 150\,mm.}
  \label{concave_prototype}
 \end{center}
\end{figure}

In this paper we show the electro-optical characterization results of one of our prototype.
This chip (Figure~\ref{concave_prototype}) is a CMV20000 global shutter CMOS image sensor from CMOSIS, with 5120$\times$3840 pixels of 6.4\,$\mu$m size. 
It has been curved with a spherical concave shape down to a radius of 150\,mm.
\section{Characterization methodology of curved CMOS}
\label{sec:methodology}
In order to know how the curving process impacts the noise properties and the performances of the detector itself, a set of characterization measurements is performed on the curved detector and on its equivalent flat version and their results are compared.
Typical measured quantities are \cite{janesick_2007,howell_2006}: detector gain, dark current, readout noise, fixed pattern noise, dynamic range and full well capacity.

The dark current is due to the thermal agitation of the electrons within the semiconductor and it is extremely sensitive to the temperature. This agitation can free the electrons which are, hence, collected in the potential wells of the pixels, becoming indistinguishable from the charges due to a direct illumination.
It is usually obtained by reading out the detector at different exposure times in complete darkness conditions.
As the averaged signal linearly grows as function of exposure time, the dark current is estimated by fitting these measurements.

The value of the gain (which determines how the amount of charge collected in each pixel will be assigned to a digital number in the image) is instead obtain by exposing the detector to uniform and constant illumination, at different exposure times.
Again the average signal, registered on the detector at each exposure time, grows linearly as function of exposure time until it reaches saturation.
More details regarding the methodologies adopted to measure all the mentioned characterization values are provided in the following subsections.

\subsection{Data acquisition}
The following measurements were performed:
\begin{itemize}
    \item For each exposure time, 30 frames were acquired.
    \item Firstly exposures to uniform light -- also called flat fields -- have been performed, and then exposures in complete darkness.
    \item The exposure time has been changed with values ranging from 0.0002\,s to saturation, for flat field exposures, and to 0.96\,s for the dark exposures.
    \item The concave ($R_\mathrm{c}=$150 mm) and a flat CMOS have been tested using the same set of measurements.
    \item The measurements were performed with camera gain set to 1.
\end{itemize}

The communication between sensor and computer was made through the CMV20000 evaluation kit.
In order to achieve a uniform illumination of the sensors, we used an integrating sphere and placed it at 1.07 m distance from the sensor.
The entrance port of the sphere was illuminated by a tungsten bulb placed inside another smaller integrating sphere.
A series of light baffles were located all around the light path to reduce the scattered light.
In order to avoid systematics due to the drifting of the light level or small ($<0.1^{o}$C) temperature fluctuations the exposure time was sampled almost randomly, by alternating measurement sets with shorter and longer exposure time.

The measurements were performed at a room temperature of 21.9$\pm0.1^o$C for the concave sensor and 21.0$\pm0.1^o$C for the flat one.
The CMOS temperature was monitored by the internal temperature sensor (calibrated for each sensor with thermocouples) and, during the full characterization, resulted to be 35.1$\pm0.2^o$C for the concave sample and $34.9\pm0.2^o$C for the flat sensor (the errors represent a variation of temperature during the measurements and not the error on the absolute value).

\subsubsection{Methodology used for dark current and readout noise measurements}
\label{sec:dark_data}
After acquiring the measurements, a series of data processing was required to obtain the desired characterization values.
As previously mentioned for each exposure time a set of 30 dark exposure frames was acquired and a median image was obtained.
By averaging further over the pixels, a mean signal level was obtained.
By fitting the linear trend of the average signal as function of exposure time one obtains the dark current and the bias level (which is a positive offset set up by the electronics to avoid recording negative numbers).

An important noise component is the temporal noise, which is the variation in time of the output value of the pixels exposed to a constant illumination.
It is typically composed of the shot noise, due to the dark current or light exposure, and of the noise generated when reading out the pixels (readout noise, RON). 
The temporal noise is obtained by computing the standard deviation of each pixel in the 30 frames acquired for each exposure time: an image is created, where each pixel is a standard deviation of the mean value of the pixel in the 30 frames.
The temporal noise (for a specific exposure time) is, hence, the mean of these standard deviations.

From the temporal noise of the shortest exposure time dataset acquired in complete darkness, we measured the RON.

\subsubsection{Methodology used for gain and PRNU measurements}
\label{sec:flat_data}
The gain and the pixel response non-uniformity (PRNU) are measured from flat field images.
As before 30 frames were acquired for each exposure time (sampling the exposure time randomly).
From these, a median image was obtained for each exposure time.
Then an average over all pixels was computed, thus providing a mean signal level.
The exposure times were sampled alternating short and long ones (as before) and they ranged between 2.4$\times10^{-4}$\,s and 0.48\,s, to allow the sensor to reach saturation.

Additionally, we estimated the temporal noise as in Section~\ref{sec:dark_data}, but this time using the flat field exposures.
The square of the temporal noise, $\sigma_{\mathrm{temp}}$, can be written as \cite{janesick_2007}:
\begin{equation}
\sigma_{\mathrm{temp}}^2 = \mathrm{const} + k(S_{\mathrm{mean}}-S_{\mathrm{offset}}),
\label{eq:temp_noise}
\end{equation}
where $k$ is the gain in units of $\mathrm{DN/e}^{-}$ and $S_{\mathrm{mean}}-S_{\mathrm{offset}}$ are the mean signal on the median image and the bias level (computed from the dark exposures) respectively.
Equation~\ref{eq:temp_noise} shows the linear relation between the mean signal and the temporal noise, which leads to the estimation of the gain, $k$, by making acquisitions at different the exposure times.

Typically the noise of an image contains also the PRNU. This is due to slightly different responses to incoming light, among the pixels of a sensor, and it is directly proportional to the number of electrons detected through the PRNU factor, $f_{\mathrm{PRNU}}$.
The total noise of an image can be written as \cite{howell_2006}:
\begin{equation}
\sigma_{\mathrm{tot}} = \sqrt{\sigma_{\mathrm{e}}^2+\sigma_{\mathrm{RON}}^2+\sigma_{\mathrm{PRNU}}^2}
\label{eq:tot_noise}
\end{equation}
where $\sigma_{\mathrm{e}}$ is the shot noise, $\sigma_{\mathrm{RON}}$ is the readout noise, and $\sigma_{\mathrm{PRNU}}$ is the PRNU noise.
As the PRNU does not depend on time, by subtracting one frame to another with equal exposure time, we eliminated its contribution from the noise equation, and we were able to estimate the shot noise.
From Equation~\ref{eq:tot_noise} we obtained $\sigma_{\mathrm{PRNU}}$ and estimated $f_{\mathrm{PRNU}}$, which is usually expressed as a percentage of the mean signal.

\section{Results}
\label{sec:results}
In Section~\ref{sec:methodology} we described the acquisition methodology and the data processing applied to obtain the fundamental values for characterizing the sensors.
In this Section we show the results obtained from the different datasets for the concave and flat samples.

\begin{figure}[ht]
 \begin{center}
  \includegraphics[width=0.98\textwidth]{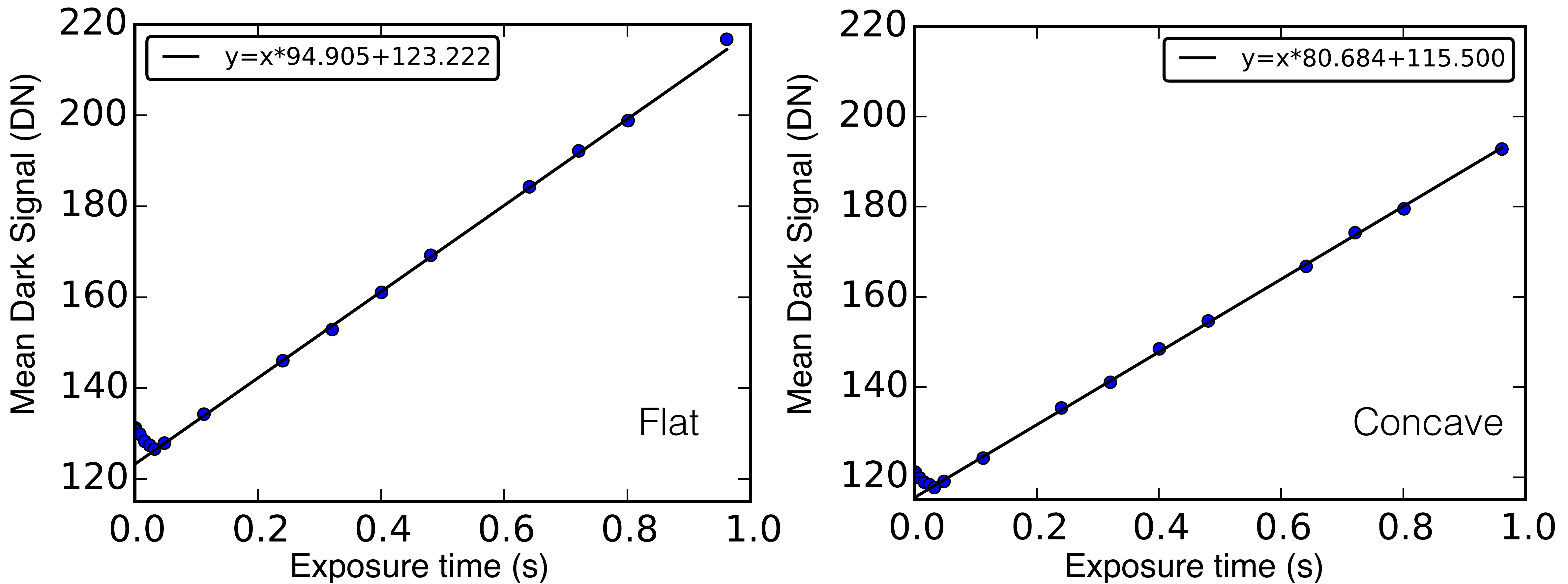}
  \caption{Mean dark exposure signal as function of exposure time. The black line is the fit to the data for exposure time larger than 0.048\,s. Left: data for the flat sensor. Right: data for the concave sensor.}
  \label{dark_current}
 \end{center}
\end{figure}

Following the prescription in Section~\ref{sec:dark_data}, we obtained the results in Figure~\ref{dark_current}.
The linear increase of measured signal for exposure times larger than 0.048\,s is due to the accumulation of charges caused by the dark current.
By fitting these data we estimated the dark current in DN/s (the slope of the fit) and the bias level (the intercept of the fit).
We obtained for the dark current a value of 80.68$\pm$0.70\,DN/s for the concave sample (at 35.1$^o$C) and of 94.90$\pm$0.59\,DN/s for the flat sensor (at 34.9$^o$C).
The errors here are the 1$\sigma$ errors on the linear fit.

For the shorter exposure times, the counts on the sensor increase and the response is not linear.
As this feature is present in both sensors, we concluded that it is an intrinsic characteristic of the CMV20000.
We applied the definition of bias level -- the mean value of the median frame with the shortest exposure time acquired in darkness -- and therefore we used this higher value, specified in Table~\ref{tab: values}, as bias level in the following.

The dark exposures also allowed us to estimate the readout noise from the temporal noise of the dataset with the shortest exposure time, as explained in Section~\ref{sec:dark_data}.
The values obtained are in Table~\ref{tab: values}.
In order to asses if the dark current and readout noises present any significant difference among the two sensors, we first evaluated their gain.
Only after converting the DN in number of electrons we can have a fair comparison.

\begin{figure}[ht]
 \begin{center}
  \includegraphics[width=0.97\textwidth]{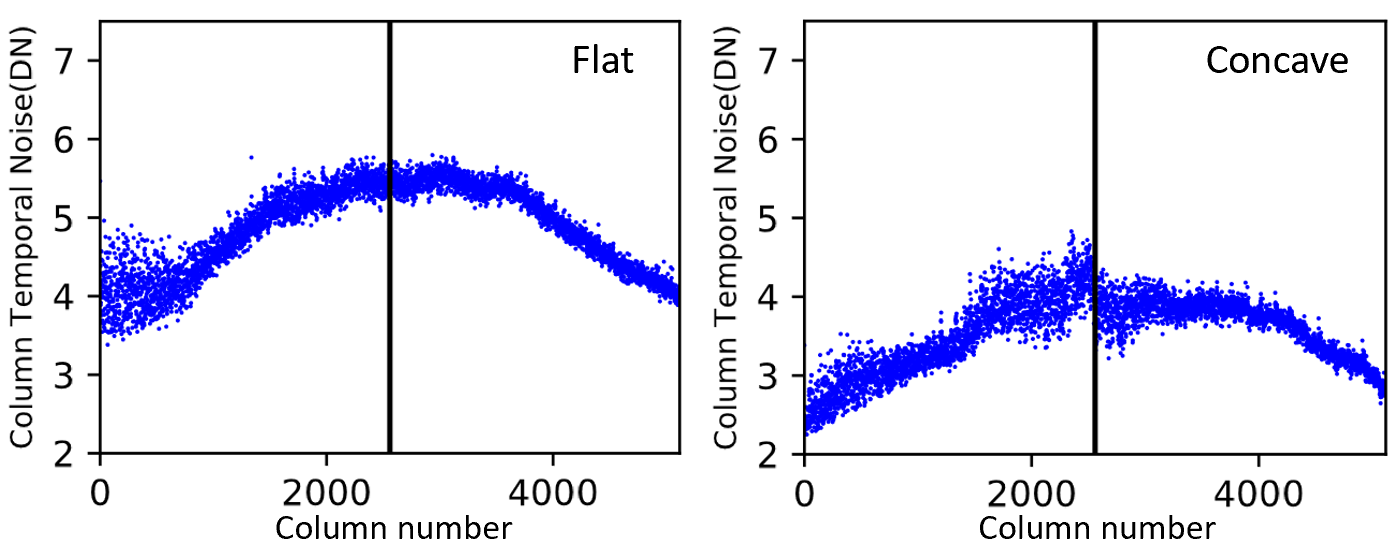}
  \caption{Column temporal noise of a median dark exposure image at 0.0002\,s vs column number for the flat (left) and the concave (right) sensors.}
  \label{column_temp}
 \end{center}
\end{figure}

Before measuring the gain, another useful diagnostic way to evaluate the performances of CMOS sensors is the column temporal noise.
This is done by plotting the standard deviation of each sensor column in the median image (from the 30 images acquired) of the dark exposure with the shortest exposure time (in this case).
The column temporal noise as function of column number is shown in Figure~\ref{column_temp} and does not present any particular difference between the two sensors.
We notice a slightly larger scatter of the values for the concave sample but such values show a similar behaviour compared to the flat sensor case -- they increase at the center and decrease at the two edges -- and they are also overall smaller.

\begin{figure}[ht]
 \begin{center}
  \includegraphics[width=0.98\textwidth]{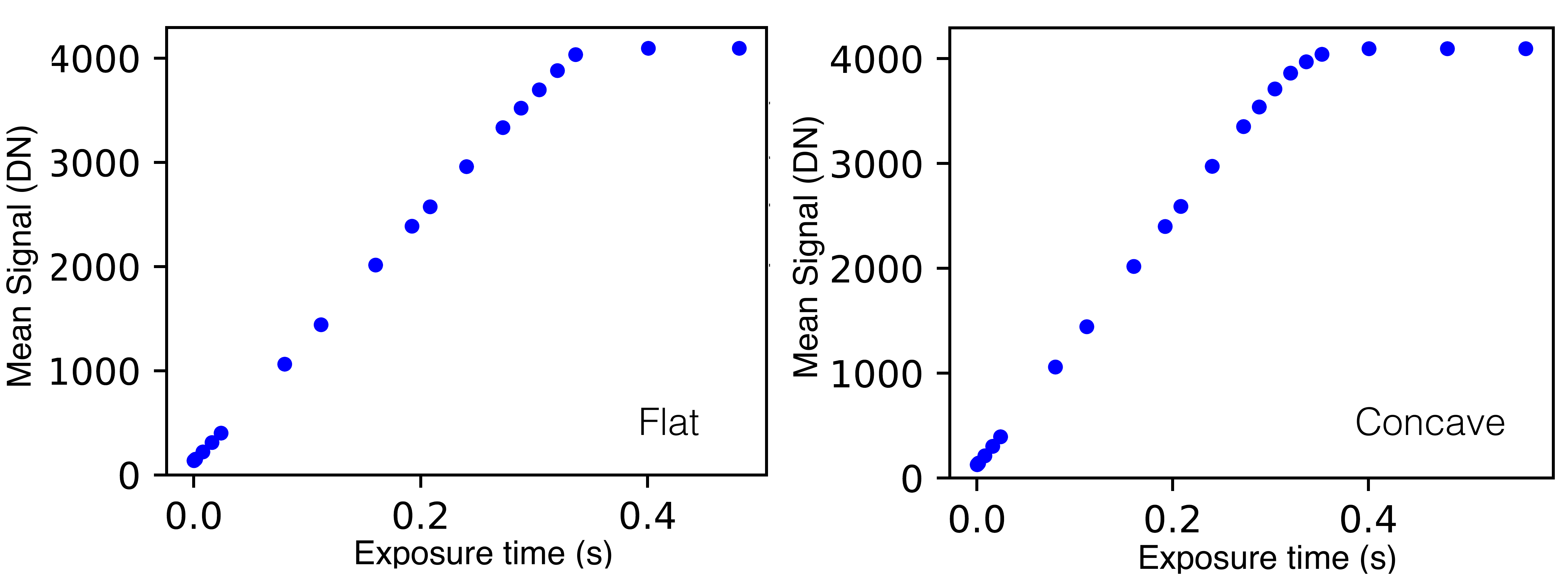}
  \caption{Mean flat field signal as function of exposure time. Left: data for the flat sensor. Right: data for the concave sensor.}
  \label{mean_f}
 \end{center}
\end{figure}

The gain was measured as explained in Section~\ref{sec:flat_data}, from a set of flat fields where the sensor is exposed to uniform and stable illumination.
In Figure~\ref{mean_f} the mean signals observed by the two sensors are shown vs the exposure times.
The signal grows linearly for exposure times larger than 0.048\,s until it reaches the saturation limit of 4095\,DN (set by the Analog Digital Converter, as it has 12-bit per pixel) for both sensors.
From this we estimated the full well capacity by subtracting the bias level to the saturation limit, and the dynamic range as $DR = 20log(S_{\mathrm{max}}/\mathrm{RON})$ where $S_{\mathrm{max}}$ is the saturation limit and RON is the readout noise.
These values are shown in Table~\ref{tab: values}.

\begin{figure}[ht]
 \begin{center}
  \includegraphics[width=0.99\textwidth]{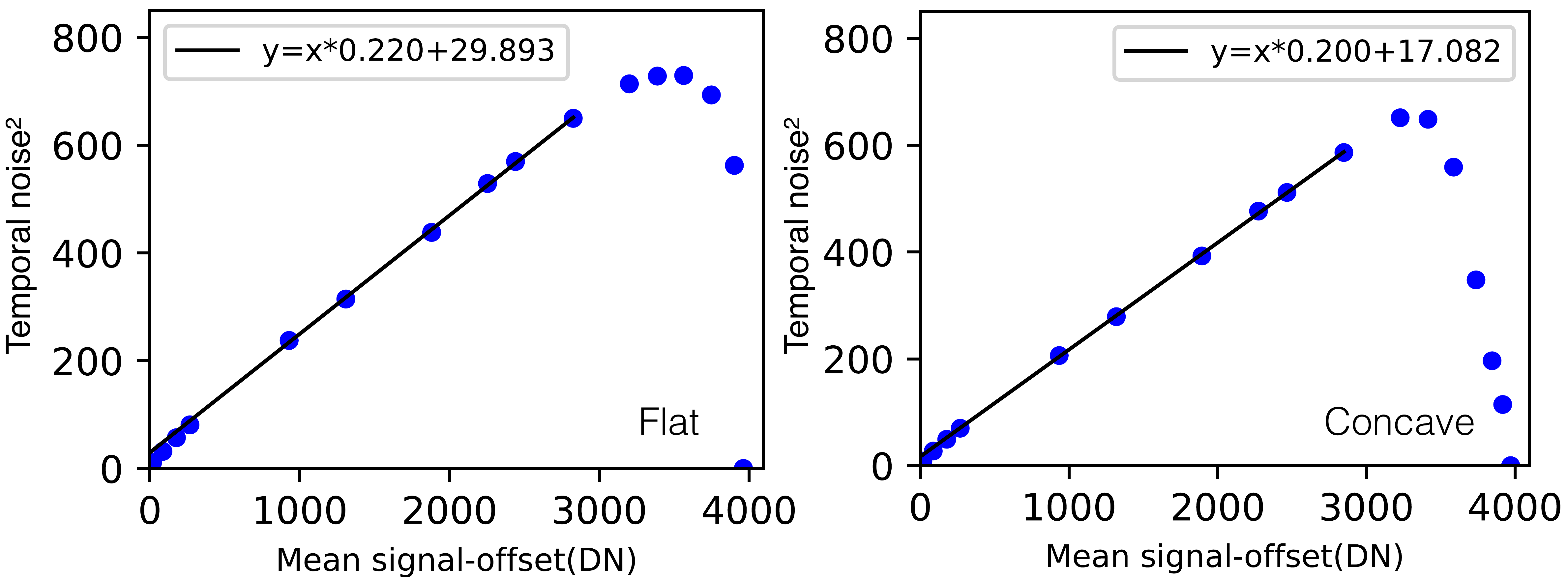}
  \caption{Squared temporal noise of flat field frames as function of mean signal-offset (bias level). The black line is the fit to the data for mean signal-offset between 1000\,DN and 3000\,DN. Left: data for the flat sensor. Right: data for the concave sensor.}
  \label{temp_no}
 \end{center}
\end{figure}

In Figure~\ref{temp_no} are plotted the squared temporal noises of both detectors against the mean signal - offset.
For values of mean signal - offset between 1000\,DN and 3000\,DN we see a linear trend as the one described by Equation~\ref{eq:temp_noise}.
From the slope of the fit to the linear part, we obtained a gain of 0.200$\pm$0.002\,DN/e$^-$ and 0.220$\pm$0.003\,DN/e$^-$ for the concave and flat sensors respectively (again, the errors are the 1$\sigma$ errors on the linear fit).
This two values are apart by 10\%, however we cannot conclude that this difference is due to the curving process as it might be within the manufacturing scatter.
The gain quoted by the manufacturer is of 0.25 DN/e$^-$ \cite{cmv20000}, 12\% larger than the gain of the flat sensor measured in this paper.
 

\begin{table}[h]
\begin{center}
\caption{Characterization values measured for the flat and concave CMV20000, with a radius of curvature $R_\mathrm{c}=$150\,mm.
Note that the same methodology has been applied to both sensors.}
\begin{tabular}{lcc}
\hline\\[-1.8ex]
   & Flat    & Concave ($R_\mathrm{c}=$150 mm)  \\
\hline\\[-1.8ex]
Bias (e$^-$)  &    595.9$\pm$24.2   & 604.0$\pm23.9$ \\[0.25ex] 
Dark current (e$^-$/s) @ 35$^o$C  &   431.39$\pm$2.7    & 403.4$\pm$3.5\\ [0.25ex]
Gain (DN/e$^-$)  & 0.220$\pm$0.003      & 0.200$\pm$0.002 \\ [0.25ex]
RON (e$^-$)  & 11      & 10 \\ [0.25ex]
Saturation (DN)  & 4095      & 4095 \\ [0.25ex]
Dynamic range (dB) & 64.74 & 66.26 \\ [0.25ex]
Full well capacity (e$^-$) & 18018 & 19871 \\ [0.25ex]
PRNU factor & 1.2\% & 2.0\% \\ [0.25ex]
\hline\\[-1.1ex]
\end{tabular}
\label{tab: values}
\end{center}
\end{table}

By applying the gains to the dark current values measured previously we find 
403.4$\pm$3.5\,e$^-$/s and 431.39$\pm$2.7 e$^-$/s for the concave and flat sensors respectively.
These two values were obtained at average temperatures ($\sim35^o$C) that differed of 0.2$^o$C.
However their temperatures can be considered the same within the errors.
We find, therefore, a smaller (of 7\%) value of dark current for the concave sample.


Finally, making use of Equation~\ref{eq:tot_noise} as mentioned in Section~\ref{sec:flat_data} we measured the PRNU factor, $f_{\mathrm{PRNU}}$, and we found it to be 2.0\% for the concave sensor and 1.2\% for the flat sensor.

\begin{figure}[ht]
 \begin{center}
  \includegraphics[width=0.99\textwidth]{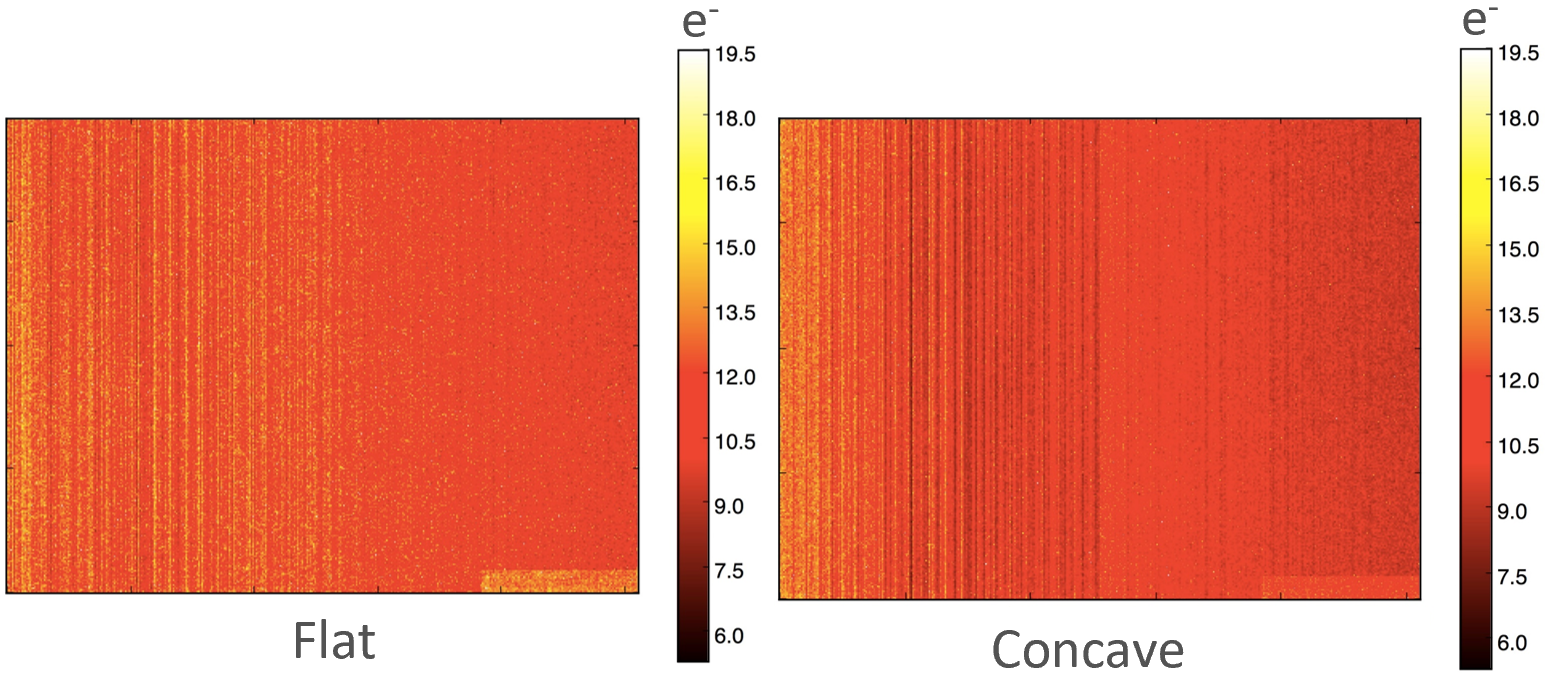}
  \caption{2D map of the RON for the flat sensor (left) and the concave one (right). The values are in numbers of electrons and are averaged in a 2$\times$2 pixels box.}
  \label{ron_map}
 \end{center}
\end{figure}

All the quantities mentioned above are averaged values across the full frame. In order to evaluate the noise behaviour of the two sensors, we also plotted the 2D map of the RON. 
This is shown in Figure~\ref{ron_map} where we averaged the pixel values in a box of 2$\times$2 to reduce the size of the images.
We see that also the 2D map does not show any particular difference among the two.
The feature in the lower right corner of the two images is an artificial effect caused by the Evaluation Bord used to control and readout the sensors.
This is generated by a bad remapping for the very short exposure times.
Note that this does not affect the results shown previously, as they consider averaged quantities computed on the full frame (hence on a large number of pixels).

\section{conclusions}
\label{sec:sections}

We presented here the characterization results of the first curved CMOS detector prototype developed within a collaboration between CNRS-LAM and CEA-LETI.
This fully-functional detector with 20\,Mpix (CMOSIS CMV20000) has been curved down to a radius of 150\,mm over a size of 24x32\,mm$^2$. 
After establishing a characterization methodology, we measured the main noise components for CMOS detectors and the gain.
We performed these measurements twice: first on the curved sensor and then on a  CMOSIS CMV20000 flat sensor.

From Table~\ref{tab: values} we have an overview of the results and we find them to be homogeneous between the flat and curved case.
By comparing the values obtained for the dark current at 35$^o$C, we see a decrease of 7\% of dark current for the concave detector.
This might be due to the curving process or, as the difference is not too significant, to intrinsic properties of the die.
We also measure a smaller readout noise of 10\,e$^-$ for the concave sensor with respect to the 11\,e$^-$ for the flat sensor.
This smaller RON generates a larger dynamic range of 66.26\,dB, against the 64.74\,dB of the flat sensor.
Also the 2D map of the RON does not show any particular difference in the noise pattern among the two sensors.

We find no significant difference in the bias level, as both values match within the errors.
We also find similar behaviour of the column temporal noise between the two sensors, where the concave sensor presented smaller values compared to the flat one.
From the measurements, the gains show a discrepancy of 10\% between each other, which might be due to an intrinsic characteristic that the chip already had before curving it. 

The PRNU factor of the concave sensor shows an increase of 0.8\% with respect to the flat sensor one.
The difference between the two is not significant.
However more investigations are required as it might be due to the curving process and it could explain the appearance of a strong 2D pattern for higher illumination levels.

From the overall performances tested in this paper, we conclude that the curving process does not impact the main characteristics of the detectors.
As more detectors are now available for testing, soon we will be able to produce a more statistical analysis on a larger sample.

\acknowledgments 
 
The authors acknowledge the support of the European Research council through the H2020 -
ERC-STG-2015 – 678777 ICARUS program. This activity was partially funded by the
French Research Agency (ANR) through the LabEx FOCUS ANR-11-LABX-0013. 

\bibliography{report} 
\bibliographystyle{spiebib} 

\end{document}